\documentclass{amsart}
\usepackage{amssymb}
\usepackage{latexsym}
\usepackage{amsmath}
\usepackage{euscript}
\usepackage{graphics}
\usepackage[all]{xy}

\newcommand{\be}{\begin{equation}}
\newcommand{\ee}{\end{equation}}
\newcommand{\ba}{\begin{eqnarray}}
\newcommand{\ea}{\end{eqnarray}}
\newcommand{\baa}{\begin{eqnarray*}}
\newcommand{\eaa}{\end{eqnarray*}}
\newcommand{\bb}{\begin{thebibliography}}
\newcommand{\eb}{\end{thebibliography}}

\newcommand{\bi}[1]{\bibitem{#1}}
\newcommand{\lab}[1]{\label{#1}}
\newcommand{\re}[1]{(\ref{#1})}

\newcounter{my}
\newcommand{\he}%
   {\stepcounter{equation}\setcounter{my}%
   {\value{equation}}\setcounter{equation}0%
   }%
\newcommand{\she}%
   {\setcounter{equation}{\value{my}}%
    }%

\renewcommand\t{\tilde}

\newcommand\ve{\varepsilon}

\newcommand\vphi{\varphi}

\newtheorem{pr}{Proposition}

\newtheorem{lem}{Lemma}

\theoremstyle{definition}

\numberwithin{equation}{section}

\begin{document}


\title{Algebraic Heun operator and band-time limiting}

\author{F.Alberto Gr\"unbaum}
\author{Luc Vinet}
\author{Alexei Zhedanov}

\address{Department of Mathematics, University of California, Berkeley
CA 94705}

\address{Centre de recherches \\ math\'ematiques,
Universit\'e de Montr\'eal, P.O. Box 6128, Centre-ville Station,
Montr\'eal (Qu\'ebec), H3C 3J7}

\address{Department of Mathematics, School of Information, Renmin University of China, Beijing 100872,CHINA}

\begin{abstract}

We introduce the algebraic Heun operator associated to any bispectral pair of operators.  We show that these operators are natural generalizations of the ordinary Heun operator. This leads to  a simple construction of the operators commuting with the projection operators in problems of band-time limiting  and it gives a way to adapt a construction first used by Perline in a purely finite setup to quite a few other situations. We also extend his algebraic construction to cover some purely finite cases where his proposal fails to give a useful commuting operator.

\end{abstract}


\maketitle

\bigskip

\section{Introduction}
\setcounter{equation}{0} 
The purpose of the present paper is twofold. First, we introduce a general algebraic Heun operator which can be obtained from a pair of bispectral operators. Second, we apply this approach to the construction of commuting differential operators in the theory of time and band limiting. This theory originated from seminal works by Slepian, Landau and Pollak on determining an unknown function supported in $[-T,T]$ from (noisy) knowledge of its Fourier transform over a band of frequencies $[-W,W]$ which involves the study of an integral operator and its eigenfunctions. Their work hinges on the existence of a differential operator commuting with the integral one. This is crucial in order to compute accurately and efficiently the eigenvectors of the integral operator, an otherwise impossible task numerically.
	
We shall expand on the observation made in \cite{GIVZ} that the generic Heun operator can be obtained by a special procedure from the hypergeometric operator and will find many new examples of either differential operators or narrow-banded matrices each  with a simple spectrum that commute with the integral operators or full matrices arising in versions of the time-band limiting problem. One can thus reduce an intractable numerical problem to a very manageable one, since the new local operators have a very spread out spectrum.

  The main result of our paper is that all known examples of such commuting operators belong to the class of algebraic Heun operators and can be explicitly  constructed by a the same  procedure. This introduction is rather brief but each section of the paper gives a fuller description of its contents.



The paper is organized as follows. In Section 2 we review the relation between the hypergeometric operator and the Heun equation (i.e. the main result of \cite{GVZ_Heun}). Section 3 can be considered as an introduction to the theory of time and band limiting. In Section 4, algebraic Heun operators are introduced through  a simple bilinear Ansatz in terms of tridiagonal operators related to the Askey scheme. In Section 5, an application of this bilinear Ansatz  to the time and band limiting of finite dimensional system is described. This goes back to \cite{Perline}. In Section 6, we consider differential operators instead of tridiagonal matrices  and study the commuting properties of these operators with projectors onto an interval. Section 7  is devoted to a reinterpretation of classical results by Slepian, Landau and Pollak in terms of algebraic Heun operators.  This is the first of the instances where the ideas in \cite{Perline} are extended to novel situations. In Section 8, we apply our approach to the case of classical orthogonal polynomials satisfying second order differential equation - i.e. to the Hermite, Laguerre and Jacobi polynomials. In Section 9, we consider an ``exceptional" case of classical orthogonal polynomial on a finite set, that cannot be treated by the simple bilinear Ansatz of \cite{Perline}. These polynomials are known as the Bannai-Ito polynomials \cite{BI}, \cite{TVZ}. We show that the reason for this phenomenon is related to the nature of the spectrum of the operators corresponding to the Bannai-Ito polynomials. In Section 10, we show how it is possible to construct an appropriate commuting operator for the time and band limiting procedure corresponding to a special case of the Bannai-Ito polynomials - the so-called anti-Krawtchouk polynomials. This operator is constructed with the help of  quadratic terms in the bispectral operators, thereby going beyond the treatment in \cite{Perline}. Finally, in a section on conclusions (Section 11), we summarize our main results and formulate some open problems.

\section{What is the Heun operator?}
	\setcounter{equation}{0}
	
The classical Heun equation  can be written  in the form \cite{Ronveaux}
\be
M f(x) = 0, \lab{Mfz} \ee 
where $M$ is a certain linear second order differential operator. It was shown in \cite{GIVZ}  
that this operator can be expressed in the following form
\be
M= \tau_1 LZ + \tau_2 ZL + \tau_3 L + \tau_4 Z + \tau_0, \lab{M_lin} \ee
where the operator $L$ is the ordinary hypergeometric operator
\be
L = x(1-x) \partial_x^2 + (\alpha_1 x + \alpha_2) \partial_x \lab{hyp_L} \ee
and $Z$ is the multiplication by the argument $x$: $Z f(x) = xf(x)$. In \re{M_lin} the parameters $\tau_i$ can be arbitrary. The operator $M$ is a second order linear differential operator such that the equation \re{Mfz} is equivalent to the generic Heun equation. It is thus natural to refer to $M$ as the Heun operator.  This method for constructing the Heun operator $M$ in terms of a bilinear combination of the operators $L$ and $Z$ is based on a procedure referred to as tridiagonalization \cite{GIVZ}, \cite{IK1}, \cite{IK2}.


The main property of the operators $L$ and $Z$ is that they form a {\it bispectral pair}. This means the following: the polynomial eigenfunctions of the hypergeometric operator $L$ are the Jacobi polynomials $P_n(x)$
\be
L P_n(x) = \lambda_n P_n(x) \lab{Jac_L} \ee
with the eigenvalues 
\be
\lambda_n = n(1+\alpha_1-n). \lab{lam_Jac} \ee
These polynomials are orthogonal and hence they also satisfy a three term recurrence relation
\be
P_{n+1}(x) + b_n P_n(x) + u_n P_{n-1}(x) = x P_n(x) \lab{3term_Jac} \ee
with certain coefficients $u_n, b_n$.


In the basis of the Jacobi polynomials the operator $L$ becomes the multiplication by $\lambda_n$ while the operator $Z$ becomes the three-diagonal operator given above. We thus see that there are two possible representations of the operators $L$ and $Z$: either $L$ is differential operator and $Z$ is a multiplication, or vice versa, $L$ is a multiplication operator while  $Z$ is a tridiagonal operator.


The key idea of the present paper is to generalize the preceding definition of the Heun operator to several bispectral situations. Once again, these situations are defined as those where a set of functions of two variables, such as $P_n(x)$ above, with x and n the variables, are eigenfunctions of two operators $L$ and $Z$ each acting on one of the variables and with eigenvalues depending solely on the other one.


In order to clarify the idea let us start with the finite-dimensional case, i.e. both operators $L$ and $Z$ can be presented by finite square matrices of the same size. We assume that these operators form a Leonard pair  \cite{Ter}.  This means  that there exist two bases $e_n$ and $d_n$ with $n=0,1,\dots,N$ such that $L$ is diagonal and $Z$ is irreducible tridiagonal in the basis $e_n$ while $Z$ is diagonal and $L$ is irreducible tridiagonal in the basis $d_n$. Tridiagonal irreducible means that all off-diagonal entries are nonzero. 


Now one can ask: what is the most general operator $M$ which has the property to be tridiagonal with respect to {\it both bases} $e_n$ and $d_n$? The answer was found by Nomura and Terwilliger \cite{NT}: this operator can be represented by  the following bilinear combination \re{M_lin}  
\be
M=\tau_1 LZ + \tau_2 ZL + \tau_3 L + \tau_4 Z + \tau_0 \mathcal{I}, \lab{Mgen} \ee
where $\tau_i$ are arbitrary coefficients and $\mathcal{I}$ is the identity operator. The sufficiency of this statement is obvious: any operator having the form \re{Mgen} is tridiagonal with respect to either basis $e_n$ or basis $d_n$ as follows from the definition of a Leonard pair. The nontrivial part of the statement is the necessity - there are no other operators with this property apart from those described by \re{M_lin}. 


In the case when $L$ is the hypergeometric operator it has already been proved  \cite{GVZ_Heun} that the resulting operator $M$ is the generic Heun operator. It is thus natural to say that for {\it any} bispectral pair $L$ and $Z$ the corresponding bilinear combination \re{M_lin} defines the algebraic Heun operator. This definition reduces the study of possible (continuous or discrete) Heun operators to the classification of basic bispectral pairs $L,Z$. In other words, any such pair $L$ and $Z$ generates a corresponding algebraic  Heun operator $M$. We thus obtain in particular an algebraic Heun operator for each entry in the Askey table. For example, starting with the Krawtchouk polynomials one can obtain a finite-dimensional Krawtchouk-Heun operator, starting from the Hahn polynomials one can obtain a Hahn-Heun operator etc. The ordinary Heun operator is thus equivalent to the Heun-Jacobi operator. Developing this perspective on algebraic Heun operators is the first motivation for the present paper. 


There is another motivation for considering such Heun operators. From the so called ``time and band limiting" problems (see details in the next sections), it is known that all explicit examples of commuting operators (for the differential case) are special cases of the Heun operator. For example in the original work on time and band limiting for the Fourier transform, the corresponding commuting operator coincides with the degenerate case of the Heun operator   that arises when the Helmholtz equation in three dimensions is separated in    prolate spheroidal coordinates \cite{Slep}.  This observation looks like a miracle and has so far remained ill-understood.  In the present paper we are going to show that a possible explanation is connected with our definition of the algebraic Heun operator. Indeed, one of the main results of our paper is that the operator of type \re{M_lin} can be taken as a corresponding commuting operator. The concrete choice of the operators $L,Z$ and $M$ will depend on the specific band-time limiting problem at hand. The basic idea in this method goes back to Perline \cite{Perline}. Our method can be considered as a generalization of Perline's Ansatz and as an embedding of this Ansatz into a generic algebraic Heun operator approach.

\section{Some background material}
	\setcounter{equation}{0}

In this paper we address a problem of wide applied interest and with this in mind, we give in this section a  review of
some background material.
	These applications range from some very old ones such as ``limited angle tomography", see \cite{G1}, to more recent ones in geodesy, see \cite{SiDa}. In all cases one runs into two kinds of ``restrictions" on an unknown function: some
apriori knowledge
on its support and the fact that one knows only a piece of its Fourier transform.
	The archetype of this situation was studied in great detail in some celebrated work of Slepian, Landau and Pollak at Bell Labs in the 1960's. For a very nice and friendly presentation of this material, see \cite{Slep,Lan}. For all the details see \cite{SLP1,SLP2,SLP3,SLP4,SLP5}.

Their work in turn was motivated by Shannon's question, see \cite{S}, in laying the mathematical foundations of Information Theory: if you have an unknown  signal
of finite duration and you are given (noisy) measurements of its Fourier transform over a limited range of frequencies, what is the best use you can make of
this information?

Phrased in mathematical terms the answer is that you should aim at recovering the projection of your unkown signal  
over the subspace of the Hilbert space spanned by a finite number of eigenfunctions of a certain integral operator. Trying anything else will suffer from the curse of numerical instability.

This raises the issue of computing these
eigenfunctions accurately and economically to assess the quality of your best possible reconstruction scheme. If you find out that this set of eigenfunctions does not span a large enough subspace to give you the resolution that you are aiming at, then you ought to measure the Fourier transform over a larger band of frequencies. This changes the integral operator whose eigenfunctions have to be computed anew to ascertain the improvement brought about
by a larger set of measurements.

The need to compute these eigenfunctions numerically is a serious bottleneck for which Slepian, Landau and Pollak found an amazing solution: they could write down analytically a second order differential operator whose eigenfunctions are automatically the eigenfunctions of the integral operator alluded to above.
The numerical computation of the eigenfunctions of this differential operator reduces the problem from a global one to a local one. Moreover the eigenvalues of the differential operator are nicely spread apart, whereas those of the integral operator are (except for a few ones) all lumped together. For the numerical aspect of this problem, look at \cite{ORX}. From a numerical point of view it is hard to imagine a better situation than that found by these Bell Labs workers. This motivates one to ask for ways to understand and extend this miracle to other scenarios.

Here is a framework, which imitates the one of Shannon and in which we can ask this type of question: suppose you have
a set of functions $\phi_n(x)$ with $n=0,1,..$ which are orthonormal with respect to a certain measure $\mu(dx)$ over a set $R$. Assume further that they are complete in the corresponding $L^2$ space. By expanding an arbitrary function $g$ as a linear combination of these $\phi_n$, we get what we refer to as the Fourier coefficients of $g$.

Consider an unknown function $f$ supported in a subset $W$ of $R$ and let $F$ stands for the generalized Fourier transform operator just described. Assume that this Fourier transform can only be known for a range $N$ of frequencies. We will argue below that our data consists of 
$$E f= \chi_N F  \chi_W f,$$
where $\chi_W$ is the {\em time limiting  operator} and  $\chi_N$ is the {\em band limiting operator}.

Above, $\chi_N$ acts by simply setting  to zero all the components with index larger than $N$ and $\chi_W$ acts by multiplication by the characteristic function of $W$. Typical choices of the set $W$ are the interval $[-W,W]$ or the interval $[0,W]$.
This terminology is borrowed from that of \cite{Slep}, and should not be taken too literally.

The apriori information on $f$ is $\chi_W f=f$ and the data is $\chi_N F f$. Combining these two pieces gives us the expression above defining $E$.  The practical solution of the problem $$E f= known$$  leads us to study the eigenvectors of the operators
$$E^*E= \chi_W F^{-1} \chi_N F  \chi_W\qquad \text{ and } \qquad E E^*= \chi_N F \chi_W F^{-1} \chi_N.$$
The operator $E^*E$, is just a finite dimensional matrix $M$, given by
\begin{equation}\label{MatrixM}
	(M)_{m,n}= \int_{-W}^W \phi_m(x)\phi^{*}_n(x) \mu(dx), \qquad 0\leq m,n \leq N 
\end{equation}
and the operator $S= E E^*$ acts by means of the integral kernel
\begin{equation}\label{kernel}
  k(x,y)=\sum_{n=0}^N \phi_n^*(x)\phi_n(y).
\end{equation}

This is entirely analogous to the situation dealt with by the Bell Labs team, or to be more accurate,
if the index $n$ runs over a continuous set as in the case of Slepian, Landau and Pollak, one has the kernel of an integral operator,
as mentioned above, namely the celebrated ``sinc kernel" 
$$ \sin(W(x-y))/(x-y) =\int_{-W}^W e^{ikx} e^{-iky} dk.$$
obtained by integrating the product $e^{ikx} e^{-iky}$ over values of $k$ in the range $[-W,W]$.
In this simple and celebrated case, since the functions $e^{ikx}$ are symmetric in $(x,k)$ the operators $E E^*$ and $E^*E$ have the same appearance. Note that the functions $e^{ikx}$ give a trivial bispectral situation, with $L,Z$ being the second derivatives (or in this very simple case even the first ones) with respect to $x$ or $k$ respectively.


Returning now to the set-up above, we ask for those situations where these full symmetric matrices (or integral operators),  would allow for a set of commuting tridiagonal ones (or second order differential operators) with a simple spectrum. This last condition on the spectrum  would guarantee that the eigenvectors of the tridiagonal matrices are automatically eigenfunctions of the full matrices in question.


We are interested in situations where we can find a commuting tridiagonal operator for 
each
value of the parameter $N$ and ``size" of the set $W$.  For the purpose of applications, the set $W$ will either go from the left end of $R$ to some place before its right end or it may be a symmetrically placed piece inside $R$.


It should be clear that we are looking for a miracle that in general will not
hold. But even if this local object exists, the question of finding it remains a
challenging problem. In this paper we will consider a number of concrete situations where the miracle holds and for which we can moreover write down explicitly the local operator in question.

\section{Algebraic Heun operator}
\setcounter{equation}{0} 
Let $L$ and $Z$ be a bispectral pair of operators. This means that there exists a set of eigenfunctions of the operator $L$
\be
L \psi(x,k) = \lambda(k) \psi(x,k) \lab{L_eig} \ee
with eigenvalues $\lambda(k)$ depending on the parameter $k$. Here it is understood that the operator $L$ acts on the space of functions of the argument $x$. The point of bispectrality is that
same functions serve as the eigenfunctions of the operator $Z$
\be
Z \psi(x,k) = \omega(x) \psi(x,k) \lab{eig_Z} \ee
with eigenvalues $\omega(x)$. It is assumed that the operator $Z$ acts on the space of functions of the argument $k$.

In the case considered by Slepian, Landau and Pollak we have 
\ba
&&L=-(d/dx)^2,  \; \lambda(k)=k^2, \nonumber \\ &&Z=-(d/dk)^2, \;  \omega(x)=x^2,\quad  \psi(x,k)=e^{ikx}. \lab{L_free_SLP} \ea
The bispectral classification problem, that is the search for all situations of the type we just described, was first considered in \cite{DG} in the case when $L$ is a second order differential operator in Schr\"odinger's form and $Z$ is a differential operator of arbitrary order. As indicated in that paper, the motivation for raising this question came from the issues and results in signal processing that we have discussed and that have been explored in several  publications, for a sample see \cite{CG5,G4,GPZ2,GY}.


The bispectrality property mentioned above can be understood as follows: 
in the $x$-representation the operator $Z$ acts as a multiplication by the function $\omega(x)$ while the operator $L$ is (typically) a differential or difference operator of the second order.
Conversely, in the $k$-representation, the operator $L$ acts as a multiplication by the function $\lambda(k)$   while the operator $Z$ is (typically) a second order differential or difference operator.


It is good to take a second look at expression \re{Mgen} and remark that in forming a product such as $L Z$ one needs to stick to one representation, either the  $x$ or the $k$ one.


The operators $L$ and $Z$ may be either differential or difference (discrete) operators. There is also an important finite-dimensional subclass of these operators. In this case the operators $L$ and $Z$ are both finite-dimensional square matrices. The case when both of them are  symmetric tridiagonal matrices as it happens for classical orthogonal polynomials on finite sets was discussed in \cite{Perlstadt}, where a construction in \cite{G} was adapted to this situation. The problem
was revisited by Perline, see \cite{Perline}, who came up with a very nice general expression for a tridiagonal matrix that commutes with the corresponding full matrix.


The main point of the present paper is to explore several situations where the construction in \cite{Perline} can de adapted beyond its original framework. We also show that, as remarked by Perline, his construction can, at times, fail to give a tridiagonal matrix with simple spectrum. This will be illustrated in an example involving the Bannai-Ito polynomials.


We display two important examples that lie beyond the situation considered in \cite{Perline}. One of the standard examples is the hypergeometric operator 
\be
L = x(1-x) \partial_x^2 + (\nu_1 x +\nu_2) \partial_x \lab{L_hyp} \ee
while $Z$ is the multiplication by $x$:
\be
Z f(x) =x f(x) \lab{Z_x} \ee
for any function $f(x)$. The set of the eigenfunctions is formed by the Jacobi polynomials
\be
L P_n^{(\alpha,\beta)}(x) = \lambda_n \: P_n^{(\alpha,\beta)}(x). \lab{L_Jac} \ee
In the dual representation, we have an infinite-dimensional vector $$\vec P =(P_0(x), P_1(x), P_2(x), \dots),$$ where the argument $x$ plays the role of a parameter. In this representation the operator $Z$ is a bispectral
tridiagonal (Jacobi) matrix. Indeed, it is well known that for any set of orthogonal polynomials (in particular, for the Jacobi polynomials) one has the 3-term recurrence relation
\be
x P_n^{(\alpha,\beta)}(x) = P_{n+1}^{(\alpha,\beta)}(x) + b_N P_n^{(\alpha,\beta)}(x) + u_n P_{n-1}^{(\alpha,\beta)}(x). \lab{3-term_Jac} \ee
The operator $L$ in this representation is the multiplication of the vector $\vec P$ by the coefficient $\lambda_n$.


Another example arises from the Heisenberg-Weyl algebra $[a, a^{\dagger}]=1$. The standard quantum-mechanical representation is
\be
a= 2^{-1/2} \left(\partial_x + x \right), \quad a^{\dagger}= 2^{-1/2} \left(-\partial_x + x \right). \lab{a_Heis_real} \ee
The operator $L$ is defined as 
\be
L = a^{\dagger} a = {1\over 2} \left(-\partial_x^2 + x^2 \right) - {1\over 2} \lab{L_osc} \ee
which coincides with the Hamiltonian of the harmonic oscillator. We define the operator $Z$ as the coordinate operator:
\be
Z = 2^{-1/2} \left( a^{\dagger} + a \right) = x .\lab{Z_Herm} \ee
The eigenfunctions of the operator $L$ are the Hermite functions
\be
L \psi_n(x) = n \psi_n(x), \lab{L_Her} \ee
where
\be
\psi_n(x) = \exp(-x^2/2) \: H_n(x) \lab{psi_Her} \ee
and $H_n(x)$ are the Hermite polynomials.


We now {\it define} the algebraic Heun operator $M$ as the generic bilinear combination of the pair of bispectral operators $L$ and $Z$:
\be
M =  \tau_1 LZ + \tau_2 ZL + \tau_3 L + \tau_4 Z + \tau_0 \mathcal{I}, \lab{M_def} \ee
where $\mathcal{I}$ is the identity operator.

The main property of the operator $M$ is that it exhibits a certain (generalized) bispectrality. In order to clarify this statement consider first the finite-dimensional case of a Leonard pair. We  assume the existence of two bases $e_n$ and $d_n$,
such that the operator $L$ is diagonal in the basis $e_n$ while the operator $Z$ is tridiagonal:
\be
L e_n = \lambda_n e_n; \quad Z e_n = \xi_{n+1} e_{n+1} + \eta_n e_n + \xi_n e_{n-1}, \quad n=0,1,2,\dots, N. \lab{LZ_e_n} \ee   
Similarly,  the operator $Z$ is diagonal in the dual basis $d_n$ while the operator $L$ is tridiagonal:
\be
Z d_n = \mu_n d_n; \quad L d_n  = a_{n+1} d_{n+1} + b_n d_n + a_n d_{n-1}, \quad n=0,1,2,\dots, N. \lab{LZ_d_n} \ee   
The only restriction on the operators $L$ and $Z$ is that the off-diagonal coefficients are nonzero: $\xi_n \zeta_n \ne 0, \; a_n c_n \ne 0, \: n=1,2,\dots$. This leads to the conclusion that the eigenvalues of $L$ and $Z$  are nondegenerate $\lambda_i \ne \lambda_j$ and $\mu_i \ne \mu_j$ if $i \ne j$.

It is then clear that the operator $M$ is tridiagonal with respect to {\it both} bases $e_n$ and $d_n$ (it was proven in \cite{NT} that the bilinear combination \re{M_def} exhausts all possible operators $M$ with this property for the finite-dimensional case).

This observation can be generalized to the infinite-dimensional case where the operators $L$ and $Z$ can be either difference or differential operators of second order.

Consider, e.g. the case when $L$ is a second order differential operator having orthogonal polynomial solutions  
\be 
L P_n(x) = \lambda_n P_n(x) , \quad n=0,1,2,\dots \lab{L_P_hyp} \ee
This means that the operator $L$ should belong to one of three types: Jacobi, Laguerre or Hermite. In this case $Z$ is the operator of multiplication by the argument $x$. Then the operator $M$ is again a differential operator of second order.    Now in the basis corresponding to the orthogonal polynomials $P_n(x)$, the operator $L$ is diagonal (multiplication by $\lambda_n$) while the operator $Z$ becomes a tridiagonal operator equivalent to the three term recurrence relation for the polynomials $P_n(x)$
\be
Z P_n(x) = x P_n(x) = P_{n+1}(x) + b_n P_n(x)+ u_n P_{n-1}(x) \lab{Z_x_P} \ee 
It is obvious, that in the basis $P_n(x)$ the operator $M$ is tridiagonal:
\be
M P_n(x) = A_n^{(+)} P_{n+1} + A_n^{(0)} P_n(x) + A_n^{(-)} P_{n-1}(x), \lab{M_P_tri} \ee
where the coefficients $A_n^{(+)},A_n^{(0)},A_n^{(i)}$ are easily expressed in terms of $u_n, b_n$ and $\lambda_n$. This construction of the operator $M$ is closely related with the tridiagonalization procedure described in \cite{Ismail}, \cite{IK1}, \cite{IK2} and developed in \cite{GVZ_Heun}. 

Finally, one can have the situation where the operator $M$ is a second-order differential operator in two different representations. One of these representations corresponds to the case when the operator $L$ is a second-order differential operator in the argument $x$ while $Z$  is a multiplication by a function $\phi(x)$
\be
Z f(x) = \phi(x) f(x) \lab{Z_f_phi} \ee
The dual representation corresponds to the choice of the operator $L$ as a multiplication by a function $\chi(k)$
\be
L f(k) = \chi(k) f(k) \lab{L_F_chi} \ee
while the operator $Z$ becomes a second order differential operator in $k$.
Roughly speaking, the algebraic Heun operator $M$ remains  generalized tridiagonal in two different representations of the operators $L$ and $Z$. By ``generalized tridiagonal" we mean either ordinary tridiagonal operators or differential operators of second order (Sturm-Liouville operators).

The name ``algebraic Heun operator" is related with the observation that in the case when $L$ is the hypergeometric operator and $Z$ is the multiplication by $x$, the equation 
\be
M f(x) = \lambda f(x) \lab{M_f_Heun} \ee
becomes the generic Heun equation \cite{GVZ_Heun}. An equivalent interpretation of this statement is the following property: the solution $f(x)$ of the Heun equation \re{M_f_Heun} can be expressed as a linear combination of Jacobi polynomials
\be
f(x) = \sum_{s=0}^{\infty} c_s P_s(x), \lab{f_lin_Jac} \ee
where the coefficients $c_s$ satisfy a three term recurrence relation (see \cite{GVZ_Heun} for details).

In contrast to the classical bispectral operators $L$ and $Z$, the algebraic Heun operator $M$ does not allow in general, an explicit solution $f(x)$ of the eigenvalue problem \re{M_f_Heun}. Indeed, it is well known, that the ordinary Heun equation, in distinction to the hypergeometric equation, cannot be ``exactly" solved   \cite{Ronveaux}. Only approximate and numerical methods can be applied to find properties of solutions of the Heun equation. Nevetheless the ``tridiagonal property" of the algebraic Heun operator $M$ still allows to apply simple algebraic methods (e.g. from the theory of orthogonal polynomials) in order to analyze solutions of the Heun equation.

Our definition of the algebraic Heun operator $M$ allows to extend the notion of Heun operator to all cases of the Askey scheme thus obtaining finite-dimensional, difference or q-difference analogs of the Heun equation. In all these cases it is natural to introduce the following nomenclature of the algebraic Heun operators: we say that the operator $M$ belongs to the ``Heun-X" type where $X$ stands for the name of the corresponding entry in the Askey scheme. We have thus introduced the Heun-Krawtchouk, the Heun-Hahn operators etc, up to the highest level of the Heun-Askey-Wilson operator.

We postpone the detailed analysis of the properties and applications of the algebraic Heun operators for a future publication. Instead, in the present paper we shall concentrate on only one (but important) application of the algebraic Heun operator $M$. Namely, we shall see that local operators of the form \re{M_def} yield operators that commute with important global operators that arise naturally in ``band-time" limiting.

\section{Band and time limiting, the finite-dimensional case}
\setcounter{equation}{0}
As explained earlier, the procedure of time-band limiting consists in the restriction of the operators $L$ and $Z$ to specific subspaces. Recall that this leads eventually to some integral operators whose eigenfunctions one needs to compute. No analytic solutions are available and accurate numerical solution of such problems is highly problematic. Nevertheless, in some cases it is possible to find differential or difference operators of second order which commute with these integral operators. This gives a numerical problem that can be handled  successfully.


The problem of constructing the corresponding tridiagonal (or second order differential) operator $T$ which commutes with these integral operators can be solved in terms of algebraic Heun operators.


In this section we demonstrate how this scheme works for the finite-dimensional case. Consider a pair of finite-dimensional operators $L$ and $Z$ defined by \re{LZ_e_n} and \re{LZ_d_n}. These operators are assumed to be irreducible, i.e. all off-diagonal entries $a_n$ and $\xi_n$ are nonzero. This means, in particular, that the spectra are nondegenerate
\be
\lambda_n \ne \lambda_m , \quad \mu_n \ne \mu_m, \quad \mbox{if} \quad n \ne m.  \lab{ll_ndeg} \ee
Let us introduce the orthonormal orthogonal polynomials $\vphi_n(x), \: n=0,1, \dots, N$ by the recurrence relation
\be
a_{n+1} \vphi_{n+1}(x) + b_n \vphi_n(x) + a_n \vphi_{n-1}(x) = x\vphi_n(x), \quad \vphi_{-1}=0,\; \vphi_0=1. \lab{rec_phi} \ee
These polynomials are orthogonal on a grid $\lambda_s$ with some weights $w_s>0$:
\be
\sum_{s=0}^N w_s \vphi_n(\lambda_s) \vphi_m(\lambda_s) = \delta_{nm}. \lab{ort_phi} \ee
The eigenvectors $e_n$ of the operator $L_n$ are expressed as
\be
e_s = \sum_{n=0}^N \sqrt{w_s} \vphi_n(\lambda_s) d_n. \lab{ed_exp} \ee
The reciprocal expansion of the basis vectors $d_n$ in terms of the $e_n$ looks as follows
\be
d_n = \sum_{s=0}^N \sqrt{w_s} \vphi_n(\lambda_s) e_s. \lab{de_exp} \ee
In a similar way we can introduce the orthonormal polynomials $\chi_n(x)$ which satisfy the recurrence relation
\be
\xi_{n+1} \chi_{n+1}(x) + \eta_n \chi_n(x) + \xi_n \chi_{n-1}(x) = x\chi_n(x), \quad \chi_{-1}=0,\; \chi_0=1. \lab{rec_chi} \ee
These polynomials are orthogonal with respect to another set of positive weights $\t w_s$:
\be
\sum_{s=0}^N \t w_s \chi_n(\mu_s) \chi_m(\mu_s) = \delta_{nm}. \lab{ort_chi} \ee
Due to the obvious duality between the operators $L$ and $Z$, we can write down two other interbasis expansions:
\be
d_s =  \sum_{n=0}^N \sqrt{\t w_s} \chi_n(\mu_s) e_n \lab{de_exp_chi} \ee
and 
\be
e_n = \sum_{s=0}^N \sqrt{\t w_s} \chi_n(\mu_s) d_s. \lab{ed_exp_chi} \ee
Relations \re{ed_exp}-\re{de_exp} and \re{de_exp_chi}-\re{ed_exp_chi} should give identical results. This is so if and only if the compatibility relation
\be 
\sqrt{w_s} \vphi_n(\lambda_s) = \sqrt{\t w_n} \chi_s(\mu_n), \quad n,s=0,1,\dots,N \lab{duality_phi_chi} \ee
holds. Relation \re{duality_phi_chi} is known as the Leonard duality condition \cite{Leonard}.

Introduce two projection operators $\pi_1$ and $\pi_2$ as follows
\be
\pi_1 e_n = \left\{ {e_n, \quad \mbox{if} \quad n \le J_1  \atop 0, \quad \mbox{if} \quad n > J_1} \right . , \quad \quad  \pi_2 d_n = \left\{ {d_n, \quad \mbox{if} \quad n \le J_2  \atop 0, \quad \mbox{if} \quad n > J_2} \right .  .  \quad \lab{pi_LM_def} \ee
The operator $\pi_1$ restricts the operator $L$ to a subspace of dimension $J_1+1$, and similarly the operator $\pi_2$ restricts the operator $Z$ to a subspace of dimension $J_2+1$. Note that obviously
\be
\pi_1^2 = \pi_1, \quad \pi_2^2 = \pi_2 \lab{sq_id} \ee
which is equivalent to the statement that both $\pi_1$ and $\pi_2$ are projection operators. Simultaneous restriction to eigensubspaces of the operators $L$ and $Z$ with these projectors leads to the following self-adjoint operators
\be
V_1 = \pi_1 \pi_2 \pi_1, \quad V_2 = \pi_2 \pi_1 \pi_2. \lab{t_LZ} \ee
The operator $\pi_1$ is not narrow-banded in the basis $d_n$ and likewise the operator $\pi_2$ is not narrow-banded  in the basis $e_n$. Hence both operators $V_1$ and $V_2$ have rather complicated nonlocal structures. In order to throw light on these, let us consider the action of the projection operator $\pi_2$ on the basis $e_n$. We have
\be
\pi_2 e_n = \pi_2 \sum_{s=0}^N \sqrt{w_s} \vphi_n(\lambda_s) d_s = \sum_{s=0}^{J_2} \sqrt{w_s} \vphi_n(\lambda_s) d_s \lab{pi2_en_1} \ee
where we used \re{ed_exp} and \re{pi_LM_def}. Expanding now the basis $d_s$ in terms of the basis $e_n$ by using \re{de_exp_chi}, we arrive at the expression
\be
\pi_2 e_n = \sum_{s=0}^{J_2} \sum_{t=0}^{N} \sqrt{w_n w_t} \vphi_s(\lambda_n) \vphi_s(\lambda_t) e_t. \lab{pi2_en_2} \ee
Now the operator $V_1=\pi_1 \pi_2 \pi_1$ acts on the basis $e_n$ as follows,
\be
V_1 e_n = \sum_{s=0}^{J_2} \sum_{t=0}^{J_1} \sqrt{w_n w_t} \vphi_s(\lambda_n) \vphi_s(\lambda_t) e_t = \sum_{t=0}^{J_1} K_{tn}^{(1)} e_t, \quad n=0,1,\dots, J_1, \lab{exp_K1} \ee
i.e. the operator $V_1$ can be represented in the basis $e_n$ by the matrix $K^{(1)}$ of dimension $(J_1+1) \times (J_1+1)$ with entries
\be
K^{(1)}_{tn} = \sum_{s=0}^{J_2} \sqrt{w_n w_t} \vphi_s(\lambda_n) \vphi_s(\lambda_t). \lab{K1_1} \ee
Using the Leonard duality relation \re{duality_phi_chi}, we can express this matrix in an equivalent form:
\be
K^{(1)}_{tn} = \sum_{s=0}^{J_2} {\t w_s} \chi_n(\mu_s) \chi_t(\mu_s). \lab{K1_2} \ee
Assume that $J_2=N$ (this means that the projection operator $\pi_2$ becomes the identity operator). Then by the orthogonality relation \re{ort_chi}, we see that
\be
K^{(1)}_{tn} = \delta_{tn} \quad t,n=0,1,\dots, J_1 \lab{K1_delta} \ee
i.e. in this case, $V_1= \pi_1$ as follows from the definition \re{t_LZ} if $\pi_2$ is the identity operator. For $J_2<N$ the operator $V_1$ is ``nonlocal" in the basis $e_n$ (i.e.{\it a priori} all entries $K_{tn}^{(1)}$ are nonzero). 

We can also express the matrix $K_{tn}^{(1)}$ in a third form using the Christoffel-Darboux identity for orthonormal polynomials 
\be
\sum_{k=0}^n \vphi_k(x) \vphi_k(y) = a_{n+1} \: \left(\frac{\vphi_{n+1}(x) \vphi_n(y) -\vphi_{n+1}(y) \vphi_n(x) }{x-y}\right). \lab{CD} \ee
With the help of \re{CD} expression \re{K1_1} can be converted to the form
\be
K^{(1)}_{tn}= \left\{ {\sqrt{w_n w_t} \: a_{J_2+1} \: \left( \frac{\vphi_{J_2+1}(\lambda_n) \vphi_{J_2}(\lambda_t) -\vphi_{J_2+1}(\lambda_t) \vphi_{J_2}(\lambda_n)}{\lambda_n-\lambda_t} \right), \quad n \ne t   \atop  a_{J_2+1} \: w_n \left( \vphi_{J_2+1}'(\lambda_n) \vphi_{J_2}(\lambda_n) - \vphi_{J_2}'(\lambda_n) \vphi_{J_2+1}(\lambda_n) \right), \quad n=t } \right.   \lab{K1_3} \ee
Note that the operator $V_1$ is symmetric:  $K^{(1)}_{tn}=K^{(1)}_{nt}$ in the basis $e_n$. By duality, the operator $V_2$ has a similar expression and we shall not write it down here.


We have thus arrived at the problem of diagonalizing of the operators $V_1$ or $V_2$. 
This is a nontrivial problem  because of the nonlocality of the  matrices representing the operators $V_1$ and $V_2$. Nevertheless, with hindsight, one can hope to find an operator $T$ which commutes with both operators $V_1$ and $V_2$. We can demand that this operator $T$  be ``as simple as possible". This means that we can try to search for a tridiagonal operator $T$ commuting with $V_1$ and $V_2$. Such a program was successfully carried out in \cite{Perlstadt} for operators belonging to the Askey scheme of classical orthogonal polynomials \cite{KLS}. Pursuing this, Perline (see \cite{Perline}) has proposed a general method for constructing the operator $T$ in the finite dimensional case. We here use the idea in \cite{Perline} to construct a wide family of commuting operators $T$.


The basic idea of the method is to choose the operator $T$ in the form of what we have called the algebraic Heun operator \re{M_def}. If the operator $M$ commutes with the projection operators 
\be
[M,\pi_1]=[M,\pi_2]=0 \lab{comm_M_pi}. \ee
then obviously, the operator $M$ commutes with the restriction operators $V_1$ and $V_2$. Thus \re{comm_M_pi} are {\it sufficient} conditions for the commutativity
\be
[M,V_1]=[M,V_2]=0 \lab{comm_M_V} \ee
and \re{comm_M_V} will be satisfied if we choose the parameters $\tau_i, \: i=0,1,\dots,4$ such that the operator $M$ satisfies conditions \re{comm_M_pi}.

First, notice the following elementary fact. 
\begin{lem}
The tridiagonal operator $M_1$ given by  
\be
M_1 e_n = A_{n+1}^{(1)} e_{n+1} + B_n^{(1)} e_n + C_n^{(1)} e_{n-1} \lab{V1} \ee 
commutes with the projector $\pi_1$ in \re{pi_LM_def} if and only if 
\be
A_{J+1}^{(1)}=C_{J+1}^{(1)}=0 \lab{AC_1=0}. \ee
Similarly, the tridiagonal operator $M_2$ given by
\be
M_2 d_n = A_{n+1}^{(2)} d_{n+1} + B_n^{(2)} d_n + C_n^{(2)} d_{n-1} \lab{V2} \ee
commutes with the projector $\pi_2$ in \re{pi_LM_def} if and only if 
\be
A_{K+1}^{(2)}=C_{K+1}^{(2)}=0. \lab{AC_2=0} \ee
\end{lem}
Consider now the operator $M$ given by \ref{M_def}. In the basis $e_n$ this operator is tridiagonal \re{V1} with the coefficients:
\ba
&&A_n^{(1)}= (\tau_1 \lambda_n + \tau_2 \lambda_{n-1} + \tau_4) \xi_n, \quad C_n^{(1)}= (\tau_1 \lambda_{n-1} + \tau_2 \lambda_{n} + \tau_4) \xi_n, \nonumber \\&& B_n^{(1)} = (\tau_1 + \tau_2) \eta_n \lambda_n + \tau_3 \lambda_n + \tau_4 \eta_n +\tau_0. \lab{ABC1} \ea
In the basis $d_n$ the operator $M$ is tridiagonal as well with the coefficients
\ba
&&A_n^{(2)}= (\tau_1 \mu_{n-1} + \tau_2 \mu_{n} + \tau_3) a_n, \quad C_n^{(2)}= (\tau_1 \mu_{n} + \tau_2 \mu_{n-1} + \tau_3) a_n, \nonumber \\ &&B_n^{(2)} = (\tau_1 + \tau_2) b_n \mu_n + \tau_3 b_n + \tau_4 \mu_n + \tau_0. \lab{ABC2} \ea
Now the conditions \re{AC_1=0} read as follows 
\be \tau_1 \lambda_{J+1} + \tau_2 \lambda_{J} + \tau_4 =0, \quad \tau_1 \lambda_{J} + \tau_2 \lambda_{J+1} + \tau_4 =0.  \lab{res_cond_1} \ee
Because of the nondegeneracy of the eigenvalues $\lambda_J \ne \lambda_{J+1}$, it follows  from \re{res_cond_1} that 
\be
\tau_2=\tau_1 \lab{tau1=tau2}. \ee
Assuming that condition \re{tau1=tau2} holds, we only have the restriction
\be
\tau_1 \left(\lambda_J + \lambda_{J+1} \right) + \tau_4 =0. \lab{res_cond_11} \ee
Similarly,  associated to the other basis,  we have the condition 
\be
\tau_1 \left(\mu_K + \mu_{K+1} \right) + \tau_3 =0. \lab{res_cond_22} \ee
From \re{res_cond_11}-\re{res_cond_22}, it follows that given $\tau_1=\tau_2$, it is always possible to choose the parameters $\tau_3$ and $\tau_4$ in order to satisfy the commutation relations \re{comm_M_pi}.

Note that the condition $\tau_1=\tau_2$ means that we can express the operator $M$ in the form
\be
M= \tau_1 \{L,Z \} + \tau_3 L + \tau_4 Z + \tau_0, \lab{M_anti} \ee
where $\{L,Z\}=LZ+ZL$ is the anticommutator. In turn, this condition is very naturally related to the Hermitian properties of the operators $L,Z$. Indeed, the operators $L$ and $Z$ are both self-adjoint in the bases $e_n$ and $d_n$. In view of that, the operator $M$ in the form \re{M_anti} will also be self-adjoint. This is basically the result obtained by Perline \cite{Perline} who started with the prescribed form \re{M_anti} for the operator $M$ in the finite-dimensional case.

Summing up, if we want the Heun operator $M$ to commute with the projectors $\pi_1$ and $\pi_2$, we should choose the restricted form of this operator that arises from \re{tau1=tau2}. It is then  always possible to choose the parameters $\tau_3$ and $\tau_4$ in order to satisfy conditions \re{comm_M_pi}.

The only caution, as already pointed out in \cite{Perline}, should be made regarding a possible ``strange" behavior of the eigenvalues $\lambda_n, \mu_n$. Indeed, suppose for instance that 
\be
\lambda_n + \lambda_{n+1}=\beta_1 (-1)^n \lab{BI_lambda} \ee
with some constant $\beta_1$.
In this case condition \re{res_cond_11} leads to a degeneration: for every even $J$ we have the same solution of equation  \re{res_cond_11}. The same is true for condition \re{res_cond_22} if the dual eigenvalues satisfy the condition
\be
\mu_n + \mu_{n+1}=\beta_2 (-1)^n \lab{BI_mu} \ee
with some constant $\beta_2$. 

Thus in the two exceptional cases corresponding to \re{BI_lambda} and \re{BI_mu} Perline's Ansatz needs to be changed. These conditions for the spectrum correspond to the case of the Bannai-Ito polynomials \cite{BI}, \cite{TVZ}. We will construct a more general commuting operator $M$ in the last section.

In all other cases corresponding to the finite-dimensional entries of the Askey tableau (e.g. Krawtchouk, Hahn, q-Racah polynomials), the algebraic Heun operator $M$ (more exactly, its special Hermitian form with the symmetric condition \re{tau1=tau2}), yields a useful solution to our problem of computing the eigenvectors of a full matrix. 

An interesting open problem is the following: is it true that every operator $T$ commuting with the restriction operators $V_1, V_2$ \re{t_LZ} should also commute with the projectors $\pi_1$ and $\pi_2$?

\vspace{10mm}

\section{Band and time limiting in case of differential operators}

Consider now the case when the operator $L$ is a differential operator of second order acting on functions defined on the real line, while $Z$ is the multiplication operator, i.e.
\be
L = a(x) \partial_x^2 + b(x) \partial_x + c(x) , \quad Z f(x) = \phi(x) f(x) \lab{L_diff} \ee
with some functions $A(x), B(x), C(x), \phi(x)$. In this case the operator $M$ will be a second order differential operator as well:
\be
M= A(x) \partial_x^2 + B(x) \partial_x  + C(x) \lab{M_diff} \ee
We wish to satisfy the commutation property
\be
[M, \pi_{\alpha}]=0 , \lab{comm_diff} \ee
where the operator $\pi_{\alpha}$ is the projection to the interval $[-\infty, \alpha]$ of the real axis $-\infty<x<\infty$. In other words, the operator $\pi_{\alpha}$ can be presented as a multiplication by the characteristic function of the interval  $[-\infty, \alpha]$:
\be
\pi_{\alpha} = 1-\theta(x-\alpha) \lab{pi_cont} \ee
where
\be
\theta(x) = \left\{  {0 \quad \mbox{if} \quad x<0 \atop 1 \quad \mbox{if} \quad x \ge 0 }  \right . \lab{Heavy} \ee
is the Heavyside function.
\begin{lem}
The operator $M$ \re{M_diff} commutes with the projection operator $\pi_{\alpha}$ if and only if
\be
A(\alpha) =0, \quad B(\alpha) = A'(\alpha). \lab{cond_alpha} \ee
\end{lem}
Similarly, if one considers the projection on the finite interval $[\alpha,\beta]$ then the projection operator $\pi_{[\alpha,\beta]}$ is defined as the multiplication by the characteristic function 
\be
\pi_{[\alpha,\beta]}= \theta(x-\beta) - \theta(x-\alpha). \lab{pi_int} \ee
Then we have:
\begin{lem}
The operator $M$ \re{M_diff} commutes with the projection operator $\pi_{[\alpha,\beta]}$ if and only if
\be
A(\alpha) =A(\beta)=0, \quad B(\alpha) = A'(\alpha), \quad B(\beta) = A'(\beta). \lab{cond_albe} \ee
\end{lem}
\noindent
The proof of these lemmas is elementary and can be found in \cite{Walter}.


Consider now the algebraic Heun operator $M$, where $L$ and $Z$ are given by \re{L_diff}. Because the parameters $\alpha$ and $\beta$ for the projection operators can be chosen independently of the operator $L$, we shall assume that the coefficients $a(x), b(x), c(x)$ do not take special values when $x=\alpha$ or $x=\beta$. We then have: 
\begin{pr}
The algebraic Heun operator $M$ commutes with the projector $\pi_{\alpha}$ if and only if
\be
\tau_2=\tau_1, \quad \tau_3 + 2 \tau_1 \phi(\alpha) =0. \lab{cond_phi} \ee
\end{pr}
\noindent
Similarly, we have: 
\begin{pr}
The algebraic Heun operator $M$ commutes with the projector $\pi_{[\alpha,\beta]}$ if and only if
\be
\tau_2=\tau_1, \quad \tau_3 + 2 \tau_1 \phi(\alpha) =0, \quad \phi(\beta) =\phi(\alpha). \lab{cond_phi_ab} \ee
\end{pr}
\noindent
As in the discrete case, the Heun operator which is useful for  time-and-band limiting can be given in the form
\be
M = \tau_1 \{L,Z\} + \tau_3 L + \tau_4 Z + \tau_0. \lab{anti_M_cont} \ee


\noindent
In the dual picture we can distinguish two possibilities:


(i) either the operator $Z$ is a differential operator while the operator $L$ is multiplication by a function. This is the ``continuous-continuous" case, and will be illustrated in the next section.




(ii) or the operator $Z$ can be represented by an (infinite) tridiagonal matrix while the operator $L$ is a diagonal matrix in the same representation. This will be called the continuous-discrete or the discrete-continuous case, and will be considered in Section 8.

In both cases the time and band limiting operators can be constructed in a manner similar to the finite-dimensional case. The only difference is that in one picture the finite-dimensional operator $V_1$ is replaced with an integral operator while in the dual picture the operator $V_2$ may be either an integral operator (in the case (i)) or a tridiagonal matrix (in the case (ii)).

In the next section we consider briefly the first possibility.

\section{The best known examples coming from signal processing and Random Matrix Theory}

We consider first the most famous example of an integral operator
that allows for a commuting differential operator, namely the case of the integral operator with the sinc kernel commuting with the 
prolate spheroidal differential operator. This case is at core of the work of 
Slepian, Landau and Pollak in the 1960's and goes along with the classical problem  of time-band limiting for the Fourier transform \cite{SLP1,SLP2,SLP3,SLP4,SLP5}.

From the point of view of quantum mechanics this is also
the simplest case, i.e. free motion in dimension one corresponding to the potential $V(x) =0$ 
in the expression for the Hamiltonian
$$H=-(d/dx)^2 +V(x).$$ 
In the notation used earlier we have 
$$L=-(d/dx)^2, \lambda(k)=k^2, Z=-(d/dk)^2, \omega(x)=x^2.$$

\bigskip

The integral operator has its kernel given by
$$ \sin(W(x-y))/(x-y) =\int_{-W}^W e^{ikx} e^{-iky} dk $$

\noindent
and the corresponding commuting differential operator acting on (an appropriate subspace
of functions in $[-T,T]$) is given by
\be
M=\partial_x (T^2-x^2) \partial_x  - W^2 x^2 .
\ee
It is a matter of a simple computation to see that this operator can be written in the Perline fashion
$$ M=  s_1 \{L,Z\} + s_2 L + s_3 Z + s_4 I $$
if one chooses
$$s_1=1/2, s_2=-T^2, s_3= -W^2, s_4=1.$$
In the expression above we follow the usual rule of replacing $Z$ by $\omega$, as remarked in section $4$. Regarding the domain of the differential operator $M$ given above, which amounts to a careful discussion of the appropriate boundary conditions, see \cite{KV}.

We now turn to the case of the radially symmetric Fourier transform, i.e. the
Bessel transform dealt with by Slepian in \cite{SLP4}.
The quantum mechanical Hamiltonian in this case (normally written with $r$ and not $x$ as the variable) is given by
$$L= -(d/dx)^2 +(\nu^2-1/4)/x^2$$
and once again the operator $Z$ or $\omega(x)$ is $x^2$.
The commuting differential operator given in \cite{SLP4} is 
\be
M=-\partial_x (G^2-x^2) \partial_x  + T^2 x^2 + G^2(\nu^2-1/4)/x^2 
\ee
and it is again possible to express this operator in the form
$$ M=  s_1 \{L,Z\} + s_2 L + s_3 Z + s_4 I $$
if one chooses
$$s_1=-1/2, s_2 =G^2, s_3= T^2, s4=(\nu^2-5/4).$$
where once again we are replacing $Z$ by $\omega(x)$.

In the work of Tracy and Widom in Random Matrix Theory, see \cite{M,TW,TWB} one finds three integral kernels obtained by considering either the ``bulk" of the spectrum and either ``hard" or ``soft" edges of it. 
The two explicit examples above are among the ones that play a role in Random Matrix Theory but had appeared earlier in the work of Slepian, Landau and Pollak. Actually the Fourier case was being considered at about the same time by these authors at Bell Labs and by Mehta at Princeton U. in the early 1960's.
The Airy case, considered by Tracy and Widom in connection with the soft edge of the spectrum, can also be seen to fit within the Perline formalism.

\section{Band and time limiting in a continuous-discrete setup: the classical orthogonal polynomials}
\setcounter{equation}{0}
In this section we consider the classical polynomials of Hermite, Laguerre and Jacobi. These polynomials are known to enjoy the ``time-band-limiting" commutation property, see \cite{G}. We show here how the commuting  local operators can be expressed in terms of the ``Perline Ansatz''.

In each case we denote by $\rho_n(x)$ the weight of the corresponding orthogonal polynomials 
\[
\begin{aligned}
&\rho(x) = \frac {e^{-x^2/2}}{\sqrt{2\pi}} \quad -\infty < x < \infty &\quad \text{(Hermite)} \\
&\rho(x) = e^{-x}x^{-\alpha} \quad 0 \le x &\quad \text{(Laguerre)} \\
&\rho(x) = (1-x)^{\alpha}(1+x)^{\beta}\quad -1 \le x \le 1 &\quad \text{(Jacobi).}
\end{aligned}
\]
Using the notation in \cite{G}, in each case we have
\[
D p_n = \Lambda_n\rho_n
\]
with
\[
D = \frac {1}{\rho(x)} \frac {d}{dx}\ \left( p(x) \frac {d}{dx} \right)
\]
and where the function $p(x)$ is given by $p(x) = \rho(x)$, $p(x) = x\rho(x)$ and $p(x) = (1-x^2)\rho(x)$ in the Hermite, Laguerre and Jacobi cases, respectively. This operator is symmetric in $L^2(\rho(x)dx)$. Furthermore, we have
\[
\begin{aligned}
\Lambda_n &= -2n &\quad \text{(Hermite)} \\
\Lambda_n &= -n &\quad \text{(Laguerre)} \\
\Lambda_n &= -n(n+\alpha+\beta+1) &\quad \text{(Jacobi)}.
\end{aligned}
\]
We first recast the explicit results in \cite{G} in terms of the ``Perline Ansatz''. Pick $N \ge 0$ and let $W$ be a point inside the support of the measure $\rho(x)$. This gives rise to the kernel
\[
k_N(x,y) \equiv \sum_0^N p_i(x)p_i(y)
\]
which acts as an integral operator in the respective intervals
\[
\begin{aligned}
(-\infty,W) &\quad \text{(Hermite)} \\
(0,W) &\quad \text{(Laguerre)} \\
(-1,W) &\quad \text{(Jacobi)}.
\end{aligned}
\]
The main result in \cite{G} states that the differential operator
\[
{\tilde D}_{W,N} = \frac {1}{\rho(x)} \frac {d}{dx} ((x-W)\rho(x) \frac {d}{dx} + A_Nx)
\]
commutes with the corresponding integral operator for an appropriate choice of the constant $A_N$, namely
\[
\begin{aligned}
A_N &= 2N &\quad \text{(Hermite)} \\
A_N &= N &\quad \text{(Laguerre)} \\
A_N &= N (N + \alpha + \beta + 2) &\quad \text{(Jacobi)}.
\end{aligned}
\]
We observe that these are all instances where the ``Perline Ansatz'' holds.

In terms of the differential operator $D$ given above and with $X$ being the multiplication by $x$,  
the following equivalent expressions hold
\[
\begin{aligned}
{\tilde D}_{W,N} &= \frac {1}{2} \{D,X\} - W D + (2N+1)X &\quad \text{(Hermite)} \\
{\tilde D}_{W,N} &= \frac {1}{2} \{D,X\} - W D + \left( \frac {\alpha+1}{2}\right)I &\quad \text{(Laguerre)} \\
{\tilde D}_{W,N} &= \frac {1}{2} \{D,X\} - W D + \left( (N+1)^2 + \left(N + \frac {1}{2}\right)(\alpha+\beta)\right)X + \frac {\alpha-\beta}{2} I &\quad \text{(Jacobi)}.
\end{aligned}
\]
In \cite{G}, one is dealing with an integral operator with kernel $k_N(x,y)$ and a differential one, ${\tilde D}_{N,W}$. By reversing the role of the ``time and frequency" variables, one could consider a full matrix $K_{N,W}$ with entries
\be
(K_{N,W})_{i,j} = \int_s^{W} \rho(x)p_i(x)p_j(x)dx
\lab{sw_lim} \ee
$0 \le i$, $j \le N$. The lower limit of integration is, in each of the three cases, the left end of the support of $\rho(x)$.

One can see that the ``Perline Ansatz'' works in this case too. The role of $D$ is taken up by the symmetric tridiagonal matrix $L$ that satisfies
\[
L \begin{pmatrix}
p_0(x) \\
p_1(x) \\
p_2(x) \\
\vdots
\end{pmatrix} = x \begin{pmatrix}
p_0(x) \\
p_1(x) \\
p_2(x) \\
\vdots
\end{pmatrix}
\]
and the role of the operator $X$ is taken up by the diagonal matrix $\Lambda$
\[
\Lambda = \left(
\begin{matrix} 
\Lambda_0 & & \\
& \Lambda_1 & \\
& & \ddots
\end{matrix}
\right).
\]
If $L_N$ and $\Lambda_N$ denote the matrices of size $(N+1) \times (N+1)$ obtained by keeping only the first $N+1$ rows and columns of $L$ and $\Lambda$, we can see that the matrix
\[
	\frac {1}{2} \{L_N,\Lambda_N\} - W \Lambda_N + \sigma_N  L_N
\]
commutes with $M_{N,W}$. The appropriate choice of the constant $\sigma_N$ is given by
\[
\begin{aligned}
\sigma_N &= 2N+1 &\quad \text{(Hermite)} \\
\sigma_N &= \frac {2N+1}{2} = (N + 1/2) &\quad \text{(Laguerre)} \\
\sigma_N &= (N+1)^2 + (N+1/2)(\alpha+\beta) &\quad \text{(Jacobi).}
\end{aligned}
\]
In all three cases above (Hermite, Laguerre and Jacobi) the corresponding commuting operator  $\t D_{W,N}$ belongs to the class of algebraic Heun operators. It is easily verified that for the Jacobi case one obtains the ordinary Heun operator while for the Laguerre and Hermite case some degenerate versions of the Heun operator arise, as expected. This can be considered as an additional justification of the nomenclature ``algebraic Heun operators". It should be noted, nevertheless, that in the Jacobi case the commuting operator $\t D_{W,N}$  is more restricted than the generic Heun operator. Indeed, the most general Heun operator is the linear combination \cite{GVZ_Heun}
\be
M = \tau_1 LZ + \tau_2 ZL +\tau_3 L + \tau_4 Z \lab{gen_M_Heun} \ee
with arbitrary parameters $\tau_i$. The commuting operator $\t D_{W,N}$ of Jacobi-type nevertheless, has the restriction 
\be
\tau_2=\tau_1 \lab{res_tau} \ee
This restriction leads to special types of the Heun equation 
\be
M \psi(x) = \lambda \psi(x). \lab{M_psi_Heun} \ee
Indeed, one can always assume that $\tau_1+\tau_2=1$ (because all coefficients $\tau_i$ are defined up to a nonessential common factor). For the special case \re{res_tau} we have $\tau_1=\tau_2=1/2$.

The generic Heun equation \re{M_psi_Heun} has the form \cite{Ronveaux}
\be
\psi''(x) + \left(\frac{\gamma}{x} + \frac{\delta}{x-1} + \frac{\ve}{x-d} \right) \psi'(x) +\frac{\alpha \beta x -q}{x(x-1)(x-d)} \psi(x) =0 \lab{red_Heun} \ee
where $\alpha, \beta, \gamma, \delta, \ve$ are parameters of the equation.

Comparing this equation with the equation arising from \re{gen_M_Heun} (when $L$ is the hypergeometric operator) we have the parameter correspondence \cite{GVZ_Heun}
\ba
&&\gamma = \nu_2, \; \delta = -\nu_1-\nu_2, \; \ve = 2\tau_2, \; d = -\tau_4 \nonumber \\ 
&&  \alpha \beta = -\tau_3 - \nu_1 \tau_2, \; q = \lambda - \tau_2 \nu_2 .   \lab{gde} \ea
It is clear that for the special case of the Heun operator where $\tau_1=\tau_2=1/2$, there is the restriction 
\be
\ve=1. \lab{ve=1} \ee
This means, for instance, that the Lam\'e equation cannot be obtained via the Ansatz with $\tau_1=\tau_2$ because for the Lam\'e case the parameters are \cite{Ronveaux}
\be
\gamma=\delta=\ve=1/2. \lab{Lame_par} \ee
Thus, the operator $M$  \re{gen_M_Heun} with $\tau_1=\tau_2$ (which corresponds to commuting operators in the band and time limiting restrictions for the Jacobi polynomials) leads to some special cases of the generic Heun equations.

The case of the Lam\'e equation was considered by Perline, see \cite{Perline1}. He observed that indeed, this ``elliptic" case is very different from its rational or trigonometric limiting cases. It is also seen to violate the necessary conditions for bispectrality discussed in \cite{DG} thus giving extra evidence of a strong
connection between bispectrality and the existence of a local operator commuting with  the global one that appears naturally.

Note finally, that the projection of classical orthogonal polynomials of type \re{sw_lim} where $N=\infty$ (i.e. there is no ``time" restriction, only the  ``band" restriction is performed) was considered recently in \cite{BO} in connection with probabilistic models.

\section{A look at Bannai-Ito polynomials}
\setcounter{equation}{0}
In this section we concentrate on a documented case of classical finite orthogonal polynomials known as the Bannai-Ito polynomials \cite{BI}.
We shall adopt the notation of  \cite{TVZ}.
These polynomials are covered by the theory developed in \cite{Perline}, but something worth noticing happens here.

The Bannai-Ito polynomials are supported on a set of $N+1$ points, and we shall
here take $N$ even. In this case the polynomials are determined by three
arbitrary positive constants, $r_1$, $e$ and $d$, see \cite{TVZ}. These polynomials, being
classical, satisfy a three term recursion relation as well as a 
three term difference equation. The tridiagonal matrices $L$ and $Z$ are symmetric. Hence there exist orthogonal matrices $T$ and $T^*$ of size $N+1 \times N+1$ that diagonalize the matrices $L$ and $Z$:
\be
 L T = T \Lambda \lab{LT_TL} \ee
and 
\be
Z T^* = T^*  \Omega \lab{ZT_TO} \ee
for appropriate diagonal matrices $\Lambda,\Omega$. The entries $T_{i,j}$ are constructed from  $p_i(x_j)$ where $x_j$ are the points where the orthogonality measure lives.

To perform our time-band limiting procedure we pick two constants $N_1,N_2$     both not larger than $N$ and form the matrices $\pi_1 T \pi_2$ and its adjoint $\pi_2 T^* \pi_1$ as well as their products $\pi_1 T  \pi_2 T^* \pi_1$ and $\pi_2 T^* \pi_1 T \pi_2$ of sizes
 $N_1 \times N_1$ and $N_2 \times N_2$ respectively.
The theory tells us that for an appropriate choice of the coefficients the matrix
\be
M= \{Z,\Lambda\} + t_1 Z + t_2 \Lambda \lab{mat_M_BI} \ee
``chopped off" to size $N_2 \times N_2$ will commute with the second of the two matrices above. The same theory tells us that for an appropriate linear combination, the matrix
\be
M= \{L,\Omega\} + s_1 L + s_2 \Omega \lab{mat_M_BI_2}, \ee
``chopped off" to size $N_1 \times N_1$ will commute with the first of these two matrices.

While these results hold true, each of the resulting tridiagonal matrices {\bf fail to have
a simple spectrum}, making them of little use for the problem of computing the
 eigenvectors of $\pi_1 T \pi_2 T^* \pi_1$ or $\pi_2 T^* \pi_1 T \pi_2$. Recall that this was the motivation for the entire enterprise.

It is interesting to point out that the phenomenon described above was anticipated by Perline in \cite{Perline} where he gives a general sufficient condition for the matrices constructed out of $(L,\Omega)$ or $(Z,\Lambda)$ to have simple spectrum. In the case of the Bannai-Ito polynomials, which were not widely known at the time when \cite{Perlstadt,Perline} were written, it is easy to see that these sufficient conditions are violated. Indeed, from the representations of the Bannai-Ito algebra \cite{TVZ}, it follows that the spectrum of the operators $L$ and $Z$ have expressions of the form
\be
\lambda_n = (-1)^n \:(n+\alpha) + \beta, \quad n=0,1, \dots, N \lab{BI_lambda_2} \ee
with parameters $\alpha, \beta$ related to parameters of the Bannai-Ito polynomials. Expression \re{BI_lambda_2} leads to some ``catastrophe" in the properties of the operator $M$.

Indeed, using concrete parameters for some finite-dimensional version of the Bannai-Ito polynomials \cite{TVZ}, we have that the diagonal matrix $\Omega$ consists of two simple interlacing sequences, giving for the entries the values $$0,-d-2,1,-d-3,2,-d-4,3,.....$$ The diagonal matrix $\Lambda$ is also made up of two simple interlacing sequences yielding the following entries $$r_1+3/2, -r_1-5/2, r_1+5/2, -r_1-7/2, r_1+7/2,.....$$

The condition in \cite{Perline} that would guarantee simple spectra requires 
looking at $$\Lambda_i+\Lambda_{i+1}$$ and being able to identify the index $i$ from this information. But, in this example, the values of this sum sequence are $$-d-2,-d-1,-d-2,-d-1,-d-2,...$$ making it impossible to achieve the task of determining the index.

Similarly, for the sequence $\Omega_i+\Omega_{i+1}$ we get  $-1, 0, -1, 0, -1, .....$.  In both cases we have an extreme violation of Perline's condition. It is remarkable that it is violated in this important example. This phenomenon follows from the general expression \re{BI_lambda_2} because
\be
\lambda_n + \lambda_{n+1} = 2 \beta -(-1)^n \lab{l+l_BI} \ee
that is, $\lambda_n + \lambda_{n+1}$ has only two different values for all $n=0,1,\dots, N$. It should be stressed, that for all the other entries of the Askey scheme (Racah, q-Racah, Hahn, etc) the spectrum has no such degeneration and the case of the Bannai-Ito is therefore the only exception.

This result, while disappointing, has a practical solution. One can produce symmetric pentadiagonal matrices with simple spectra that commute with the matrices of interest. We present an illustration of this construction in the next section.

\section{The anti-Krawtchouk polynomials}

Here we consider another example where we also have the phenomenon that we just presented with the Bannai-Ito polynomials. In this situation we can go {\bf one step furher}
and produce an explicit expression for a pentadiagonal matrix that commutes with the appropriate time-band limited matrix and has moreover a simple spectrum.
One can prove that no tridiagonal matrix can have this property.

This situation occurs for the so-called anti-Krawtchouk that have been discussed in \cite{GVYZ}. These polynomials can be considered as just one of the simplest special cases of the Bannai-Ito polynomials \cite{GVYZ}. Let us simply recall that we have a measure that lives on points $$x_i, i=0,1,...,N.$$ There is a tridiagonal matrix $L_1$ that has the anti-Krawtchouk polynomials as its eigenfunctions and a diagonal matrix $\Lambda=L_2$ which arises from a three term difference equation satisfied by these polynomials and which happens to coincide with the diagonal matrix of the three term recursion relation.

Algebraically, the anti-Krawtchouk polynomials arise through representations of the anti-spin algebra with three generators $L_1, L_2, L_3$ satisfying the commutation relations \cite{GVYZ}
\be
\{L_i, L_j\} = \ve_{ijk} L_k, \quad i,j,k=1,2,3, \lab{anti_spin} \ee
where $\{X,Y\}=XY+YX$ stands for the anticommutator and where $\ve_{ijk}$ is the totally antisymmetric tensor taking values $1, -1$ depending on parity of the triple $i,j,k$ (and which vanishes if there are coinciding entries). Relations \re{anti_spin} resemble the commutation relations of the Lie algebra $su(2)$ with the only difference that all the commutators are replaced by anticommutators.

Consider an irreducible finite-dimensional representation of the anti-spin algebra \cite{GVYZ}. Let $e_n$ be the basis in which the matrix $L_2$ is diagonal:
\be
L_2 e_n = x_n e_n , \quad n=0,1,\dots,N  \lab{anti_Lambda} \ee
and where the eigenvalues $x_n$ (i.e. the grid points) are
\be
x_n=(-1)^n (n+1/2). \lab{x_anti} \ee
The matrix $L_1$ is symmetric and tridiagonal in the canonical basis $e_n$ 
\be
L_1 e_n = a_{n+1} e_{n+1} + b_n e_n + a_n e_{n-1} \lab{anti_L} \ee
with coefficients given by, see \cite{GVYZ},
\ba
&&a_n^2 = \frac{(N+1)^2-n^2}{4}, \quad n=1,2,\dots, N, \nonumber \\
&& b_0 =(-1)^N (N+1)/2, \; b_n=0, \; n=1,2,\dots,N. \lab{ab_anti} \ea 
Note that the anti-Krawtchouk polynomials are self-dual \cite{GVYZ}. This means in particular, that the operators $L_1$ and $L_2$ have the same spectrum $x_i$ given by \re{x_anti}. Moreover, there exists the dual basis $d_n$ in which the operator $L_1$ is diagonal while the operator $L_2$ is tridiagonal
\be
L_1 d_n = x_n d_n, \quad L_2 d_n =  a_{n+1} d_{n+1} + b_n d_n + a_n d_{n-1} \lab{dual_basi_anti} \ee
with the same expressions as above for $a_n, b_n, x_n$.


The anti-spin algebra has the following Casimir operator
\be
Q=L_1^2 + L_2^2 + L_3^2 = L_1^2 + L_2^2 + (\{L_1,L_2\})^2  \lab{Q_anti} \ee 
which commutes with all the generators 
\be
[Q,L_i]=0, \; i=1,2,3. \lab{Q_anti_com} \ee
Given the $(N+1)$-dimensional representation \re{x_anti}-\re{anti_L}, the Casimir takes the value
\be
Q = (N+1/2)(N+3/2). \lab{Q_anti_N} \ee
Because the anti-Krawtchouk polynomials belong to the Bannai-Ito family of  polynomials, the bilinear Perline's Ansatz 
\be
M = \tau_1 L_1 L_2 + \tau_2 L_2 L_1 + \tau_3 L_1 + \tau_4 L_2 \lab{M_Per_anti} \ee
leads to an operator $M$ commuting with the time and band limiting operator, but with a {\bf degenerate spectrum.}


It is therefore natural to try to construct $M$ as a polynomial of higher degree in the operators $L_1, L_2$ in order to remedy this situation. If we expect the operator $M$ to be pentadiagonal with respect to the bases $e_n$ and $d_n$ we should restrict ourselves to taking expressions where the powers of the operators $L_1$ and $L_2$ do not exceed 2.

If we denote by $N_1$ and $N_2$ the parameters specifying the time and band limiting operations respectively, we can see that the following operator
\ba
&&M= \{ L_1^2 , L_2^2\} + \alpha_1 \{L_1^2, L_2\} +\alpha_2 \{ L_2^2, L_1\} + \nonumber \\
&& \alpha_3 L_1^2 + \alpha_4 L_2^2 + \alpha_5 L_1 + \alpha_6 L_2 \lab{M_anti_bi} \ea
will do the job once the coefficients $\alpha_i$ are appropriately determined.

In order to find these coefficients, we observe that the operator $M$ is a five-diagonal symmetric matrix which acts on the basis $e_n$ according to
\be
M e_n = G_n e_{n-2} + F_n e_{n-1} + H_n e_n + F_{n+1} e_{n+1} + G_{n+2} e_{n+2}, \lab{M_e_anti} \ee
where
\be
G_{n}=a_n a_{n-1} \left( x_n^2+x_{n-2}^2 +\alpha_1( x_n+x_{n-2}) +\alpha_3 \right), \quad n \ge 2 \lab{G_anti} \ee
and 
\be
F_n = a_n \left( \alpha_2 (x_n^2+x_{n-1}^2) + \alpha_5 \right), \quad n \ge 2 \lab{F_anti} \ee
(the explicit expression of the diagonal coefficient $H_n$ in \re{M_e_anti} is not needed for our purposes).

Let $\pi_{N_1}$ be the projection operator defined by
\be
\pi_{N_1} e_n = \left\{ {e_n, \quad n\le N_1 \atop 0, \quad n>N_1} \right. . \lab{pi1_anti} \ee
The operator $M$ commutes with $\pi_{N_1}$ if and only if
\be
G_{N_1+1}=G_{N_1+2}=F_{N_1+1} =0. \lab{cond1_anti} \ee

Using the explicit expressions \re{G_anti}-\re{F_anti}, we have the following three equivalent conditions
\ba
&& x_{N_1}^2+x_{N_1+2}^2 +\alpha_1( x_{N_1}+x_{N_1+2}) +\alpha_3=0, \nonumber \\ &&x_{N_1-1}^2+x_{N_1+1}^2 +\alpha_1( x_{N_1-1}+x_{N_1+1}) +\alpha_3=0, \lab{conds1_anti} \\
&& \alpha_2 \left(x_{N_1}^2+x_{N_1+1}^2 \right) + \alpha_5 =0.  \nonumber \ea
Consider now under what circumstances will $[M,\pi_{N_2}] =0$ with $\pi_{N_2}$ the projector defined by
\be
\pi_{N_2} d_n = \left\{ {d_n, \quad n \le N_2 \atop 0, \quad n>N_2} \right. .\lab{pi2_anti} \ee
Because the operators $L_1$ and $L_2$ have the same spectrum $x_n$ and the same coefficients $a_n, b_n$ in their three-diagonal representation, it is sufficient to note that in the basis $d_n$ the operator $M$ looks similar to \re{M_e_anti}. The explicit expressions of the coefficients $G_n, F_n$ in the basis $d_n$ are obtained from the corresponding expressions \re{G_anti}-\re{F_anti} by the simple transpositions $\alpha_1 \rightleftarrows \alpha_2$, $\alpha_3 \rightleftarrows \alpha_4$ and $\alpha_5 \rightleftarrows \alpha_6$. 

We thus find another set of conditions similar to \re{conds1_anti}:
\ba
&& x_{N_2}^2+x_{N_2+2}^2 +\alpha_2( x_{N_2}+x_{N_2+2}) +\alpha_4=0, \nonumber \\ &&x_{N_2-1}^2+x_{N_2+1}^2 +\alpha_2( x_{N_2-1}+x_{N_2+1}) +\alpha_4=0, \lab{conds2_anti} \\
&& \alpha_1 \left(x_{N_2}^2+x_{N_2+1}^2 \right) + \alpha_6 =0 . \nonumber \ea
We can now solve the system consisting of the six linear equations \re{conds1_anti} -\re{conds2_anti} with respect to the six unknowns $\alpha_i, \: i=1,2,\dots,6$. The solution is unique and given by: 
\ba
&&\alpha_1=(-1)^{N_1}, \; \alpha_2 =(-1)^{N_2}, \; \alpha_3 = -1-\kappa_1, \nonumber \\
&&\alpha_4 = -1-\kappa_2, \; \alpha_5 = (-1)^{N_2} \kappa_1, \; \alpha_6 =(-1)^{N_1} \kappa_2 \lab{alpha_anti} \ea
where
\be
\kappa_1 = 2{N_1}^2 + 4N_1 +5/2, \quad \kappa_2 = 2{N_2}^2 + 4N_2 +5/2. \lab{kap12_anti} \ee
This settles in a definite way the band  and time limit problem for the \\ anti-Krawtchouk polynomials. A simple check shows that the spectrum of this operator $M$ is (in general) {\bf nondegenerate}. The simplest commuting operator $M$ that is of practical  use hence requires a polynomial of degree 2 with respect to both operators $L_1$ and $L_2$. As was already mentioned, the Perline's type Ansatz with bilinear polynomial in $L_1$ and $L_2$ leads to a degeneration of the spectrum of the resulting operator $M$.  

A remark should be made concerning the choice \re{M_anti_bi} of the operator $M$. In general, such an operator should include all symmetric terms of total degree 4,3,2 and 1. All possible symmetric terms of degree four (and hence of degree two with respect to both operators $L_1$ and $L_2$) are
\be
\{L_1^2, L_2^2\}, \; (L_1L_2)^2 + (L_2L_1)^2, \; L_1 L_2^2 L_1, \; L_2 L_1^2 L_2. \lab{sym_4} \ee
It is clear however  from the defining relations of the algebra  \re{anti_spin} and the Casimir operator \re{Q_anti}, that among these four terms only one is linearly independent. All other terms can be  expressed in terms of total degree three and lower. We have chosen the term $\{L_1^2, L_2^2\}$ because of its symmetric form with respect to the transposition $L_1 \rightleftarrows  L_2$.

\noindent
All possible symmetric cubic terms are
\be
\{L_1^2, L_2\}, \; \{L_2^2, L_1\}, \; L_1L_2L_1, \; L_2 L_1 L_2 . \lab{sym_3} \ee 
Again, from the relations \re{anti_spin} it follows that only two terms among these four are independent. We can choose these independent terms as the anticommutators $\{L_1^2, L_2\}$ and  $\{L_2^2, L_1\}$.

Finally, there are three possible symmetric terms of total degree 2:
\be
\{ L_1, L_2\}, \; L_1^2, \; L_2^2 . \lab{sym_2} \ee
It follows from the expression \re{Q_anti} of the Casimir element that the square of the anticommutator $\{ L_1, L_2\}$ is expressible as a linear combination of $L_1^2$ and $L_2^2$. This means that we can restrict ourselves to the quadratic terms $L_1^2$ and $L_2^2$.

The fact that we have some linear dependence allows one to propose other commuting operators instead of the $M$ given above. One such choice is the polynomial
\ba
&&M= 2 L_2  L_1^2  L_2 +(-1)^{N_1+1} \{L_2, L_1^2\} +  (-1)^{N_2+1} \{L_2^2, L_1\}  \nonumber \\
&& (2N_2+1)(2N_2+3)/2  L_2^2  + (-1)^{N_1} ((2N_2+1)(2N_2+3)/2 +1) L_2 - \nonumber \\
 &&(4N_1^2+8N_1-1)/2 L_1^2 + (-1)^{N_2} ((4N_1^2+8N_1-1)/2 +3) L_1. \lab{alt_M_anti} \ea
This candidate has, once again, a nondegenerate spectrum.

\bigskip

\section{Conclusions}
\setcounter{equation}{0}
The main results obtained in the present paper are:
 

 (i) An algebraic Heun operator can be introduced for any bispectral pair of operators $L$ and $Z$ as an operator that generalizes the ordinary Heun operator.
 
 
(ii) It has been seen that the bilinear Perline's Ansatz provides a construction of the commuting operator $M$ for all entries of the Askey table apart from the case of the Bannai-Ito polynomials (which are not usually included into the Askey classification).


(iii) In a special case of the Bannai-Ito class, namely, for the anti-Krawtchouk polynomials, we have shown that certain pentadiagonal operators built from terms of degree no higher than two in the basic operators, provide commuting operators for the band-time limiting with simple spectra.


(iv) We have indicated  that Perline's operator is a special case of algebraic Heun operator. In particular, it was observed that it coincides with the ordinary Heun operator (with particular sets of parameters) in the case of the band and time limiting of the Jacobi polynomials. This can be seen as an explanation of the previously known fact that all commuting operators in the time and band limiting procedure are related to the Heun equation or its degenerate forms.


As a conjecture we can suggest that all possible commuting operators in the time and band limiting scheme can be constructed as a generalization of Perline's  Ansatz. This would mean that any such commuting operator (with a simple spectrum) would be a symmetric polynomials in two operators $L$ and $Z$ that form a bispectral pair. Proving this remains as a challenge.

\bigskip\bigskip
{\Large\bf Acknowledgments}
\bigskip

F. Alberto Gr\"unbaum and Alexei Zhedanov  are thankful to the Centre de Recher-ches Math\'ematiques (CRM) in Montr\'eal, Canada, for hospitality while this project was being pursued. The work of L. Vinet is supported in part by the Natural Sciences and Engineering Research Council (NSERC) of Canada. The work of F.A. Gr\"unbaum is partially supported by AFOSR, through FA9550-16-1-0175. The three authors are thankful to Prof. Satoshi Tsujimoto, Graduate School of Informatics, Kyoto University for making it possible to make progress on this paper during a visit to Kyoto.

\newpage

\bb{99}

\bi{AW} R. Askey and J.A. Wilson,  {\it Some basic hypergeometric orthogonal polynomials that generalize Jacobi polynomials}. Memoirs of the American Mathematical Society {\bf 319}, Providence, Rhode Island, 1985.

\bi{BO} A.Borodin and G.Olshanski, {\it The ASEP and determinantal point processes}, arXiv:1608.01564.

\bibitem{CG5} Castro, M.,  Gr\"unbaum, F. A.,  {\it The Darboux process and time-and-band limiting for matrix orthogonal polynomials}.  Linear Algebra Appl.  487 (2015),  328-341.


\bi{BI} E. Bannai and T. Ito, {\it Algebraic Combinatorics I: Association Schemes}. 1984. Benjamin \& Cummings,
Mento Park, CA.



\bi{DG} Duistermaat, J. J. and Gr\"unbaum, F. A., {\em Differential
equations in the spectral parameter}, Commun. Math. Phys. {\bf 103} (1986),
177--240.

\bi{GVYZ} V. Genest, L. Vinet, Guo-Fu Yu and A.Zhedanov, {\it Supersymmetry of the quantum rotor},  arXiv:1607.06967v1.



\bi{GIVZ} V. Genest, M.E.H. Ismail, L. Vinet, A. Zhedanov {\it Tridiagonalization of the hypergeometric operator and the Racah-Wilson algebra},  Proc. Amer. Math. Soc. {\bf 144} (2016), 4441--4454,  arXiv:1506.07803.


\bi{G} F.A.Gr\"unbaum, {\it A new property of reproducing kernels for classical orthogonal polynomials}, J.Math.Anal.Appl.
{\bf 95}, (1983), 491Â-500.

\bi{G1} F.A. Gr\"unbaum, {The limited angle reconstruction problem in computed tomography}, Proc. Symp. Applied Math., {\bf27} , AMS, L. Shepp editor, (1982), 43--61.

\bi{GVZ_Heun} F.A.Gr\"unbaum, L.Vinet and A.Zhedanov, {Tridiagonalization and the Heun equation}, J.Math.Physics {\bf 58}, 031703 (2017), arXiv:1602.04840.

\bibitem{G4}  Gr\"unbaum F. A., {\em Band-time-band limiting integral
operators
and commuting differential operators}, Algebra i Analiz {\bf 8} (1996),
122--126.

\bibitem{GPZ2}  Gr\"unbaum, F. A.,   Pacharoni I.,  Zurrian, I., {\it Time and band limiting for matrix valued functions: an integral and a commuting differential operator}, arXiv:1604.06510, (2016), Inverse Problems, {\bf 33} 2017.

\bibitem{GY} Gr\"unbaum F. A., Yakimov M., {\em The prolate spheroidal phenomenon as a consequence of bispectrality}.  Superintegrability in classical and quantum systems, CRM Proc. Lecture Notes, vol {\bf 37}, Amer. Math. Soc., Providence, RI, 2004,  301--312.

\bi{Ismail} M.E.H.Ismail, {\it Classical and Quantum orthogonal polynomials in one variable}.
Encyclopedia of Mathematics and its Applications (No. 98), Cambridge, 2005.

\bi{IK1} M. E. H. Ismail and E. Koelink. {\it Spectral analysis of certain Schr\"odinger operators}. SIGMA,
{\bf 8}: 61Â-79, 2012.

\bi{IK2} M. E. H. Ismail and E. Koelink. {\it The J-matrix method}.
Adv.Appl.Math., {\bf 56}, 379--395, 2011.


\bibitem{KV}  Katsnelson V., {\em Selfadjoint boundary conditions for the prolate spheroidal differential operator}, arXiv:1603.07542, (2016).

\bibitem{KLS} R. Koekoek, P.A. Lesky, and R.F. Swarttouw. {\it Hypergeometric orthogonal polynomials and their q-analogues}. Springer, 1-st edition, 2010.

\bibitem{Leonard}  D.Leonard, {\it Orthogonal Polynomials, Duality and Association Schemes}, SIAM J. Math. Anal., {\bf 13} (1982), 656-663.



\bi{M} M. L. Mehta, {\em Random Matrices}, 3nd ed., Elsevier Inc.
 2004.


\bi{NT} K.Nomura and P.Terwilliger, {\it Linear transformations that are tridiagonal with respect to both eigenbases of a Leonard pair}, Lin.Alg.Appl.
{\bf 420} (2007), 198--207.  arXiv:math/0605316.

\bibitem{ORX}  Osipov A.,  Rokhlin V., Xiao H., Prolate spheroidal wave functions of order zero, Mathematical tools for bandlimited approximation, Springer 2014.

\bi{Perline}  R.K.Perline, {\it Discrete Time-Band Limiting Operators and Commuting Tridiagonal Matrices}, SIAM. J. on Algebraic and Discrete Methods, 8(2), 192-Â195 (1987).

\bi{Perline1} R.K. Perline, {\it Time-band limiting matrices and Lame's equation} Numerical functional analysis and Optimization, 9 (11-12), pp 1139--1175, (1987-88)

\bi{Perlstadt} M.Perlstadt, {\it A Property of Orthogonal Polynomial Families with Polynomial Duals}, SIAM J. Math. Anal., {\bf 15}(5), 1043Â-1054 (1984).

\bibitem{Ronveaux} A. Ronveaux (Ed.), {\it Heun's Differential Equations}, Oxford University Press, Oxford, 1995. 

\bi{Roos} O. Roos, {\it Position-dependent effective masses in semiconductor theory}, Phys. Rev. {\bf B 27} (1983), 7547--7552.

\bibitem{S}  Shannon C., {\em A mathematical theory of communication}, Bell Tech. J., vol {\bf 27}, 1948, 379--423 (July) and  623--656 (Oct).

\bi{SiDa} F.J. Simons, F. A. Dahlen,  {\it Spherical Slepian functions on the polar cap in geodesy} Geophys. J. Int., 166, 1039--1062, 2006.

\bi{SLP1} D. Slepian and H. O. Pollak, {\em Prolate spheroidal wave
functions, Fourier Analysis and Uncertainty, I}, Bell System Tech. Journal,
Vol.~40,
No.~1 (1961), 43--64.

\bi{SLP2} H. J. Landau and H. O. Pollak, {\em Prolate spheroidal
wave
functions, Fourier Analysis and Uncertainty, II}, Bell System Tech. Journal,
Vol.~40, No.~1 (1961), 65--84.

\bi{SLP3} H. J. Landau and H. O. Pollak, {\em Prolate spheroical
wave
functions, Fourier Analysis and Uncertainty, III}, Bell System Tech. Journal,
Vol.~41, No.~4 (1962), 1295--1336.

\bi{SLP4} D. Slepian, {\em Prolate spheroidal wave functions,
Fourier
Analysis and Uncertainty, IV}, Bell System Tech. Journal, Vol.~43, No.~6
(1964),
3009--3058.

\bi{SLP5} D. Slepian, {\em Prolate spheroidal wave functions, Fourier
Analysis
and Uncertainty, V}, Bell System Tech. Journal, Vol.~57, No.~5 (1978),
1371--1430.

\bi{Slep} D. Slepian, {\em Some comments on Fourier analysis, uncertainty and modeling} , SIAM Review, vol 25, 3 , July 1983, pp 379--393.

\bi{Lan} H.Landau, {\em An overview on time and frequency limiting} J.F. Price (editor), Fourier Techniques and Applications, Plenum Press, New York 1985. pp 201--220.

\bi{Ter} P.Terwilliger, {\it Two linear transformations each tridiagonal with respect to an eigenbasis of the other}, Lin.Alg.Appl. {\bf 330} (2001), 149Â-203.

\bi{TVZ} S.Tsujimoto, L.Vinet and A.Zhedanov {\it Dunkl shift operators and Bannai-Ito polynomials}, Advances in Mathematics
{\bf 229}, (2012), 2123-Â2158.

\bi{TW} C. A. Tracy and H. Widom, {\em Level spacing distribution
and the
Airy kernel}, Phys. Lett. B {\bf 305} (1993), 115--118.

\bi{TWB} C. A. Tracy and H. Widom, {\em Level spacing distribution
and the
Bessel kernel}, Commun. Math. Phys. 161, (1994) pp 289--309

\bi{Walter} G.Walter, {\it Differential operators which commute with characteristic functions with applications to a lucky accident}, Complex Variables (1992), {\bf 18},  7--12.

\eb

\end{document}